\documentclass[
  aps,
  prd,
  nofootinbib,
  amsmath,
  amsfonts,
  preprintnumbers,
  showpacs,
  english
]{revtex4-2}

\usepackage{babel}
\usepackage{amsmath,amssymb,amsthm}
\usepackage{mathrsfs}
\usepackage{bm}
\usepackage{cancel}
\usepackage{relsize}
\usepackage{exscale}
\usepackage{scalerel}

\usepackage{graphicx}
\graphicspath{{figures/}}

\usepackage{booktabs}
\usepackage{array}
\usepackage{multirow}
\usepackage{dcolumn}
\usepackage{float}
\usepackage{wrapfig}
\usepackage[figtopcap]{subfigure}

\usepackage{xcolor}
\colorlet{BLUE}{blue}
\usepackage{microtype}

\usepackage{tikz}
\usetikzlibrary{svg.path}

\definecolor{orcidlogocol}{HTML}{A6CE39}

\tikzset{
  orcidlogo/.pic={
    \fill[orcidlogocol] svg{
      M256,128c0,70.7-57.3,128-128,128
      C57.3,256,0,198.7,0,128
      C0,57.3,57.3,0,128,0
      C198.7,0,256,57.3,256,128z
    };
    \fill[white]
      svg{
        M86.3,186.2H70.9V79.1h15.4v48.4V186.2z
      }
      svg{
        M108.9,79.1h41.6c39.6,0,57,28.3,57,53.6
        c0,27.5-21.5,53.6-56.8,53.6h-41.8V79.1z
        M124.3,172.4h24.5c34.9,0,42.9-26.5,
        42.9-39.7c0-21.5-13.7-39.7-43.7-39.7
        h-23.7V172.4z
      }
      svg{
        M88.7,56.8c0,5.5-4.5,10.1-10.1,10.1
        c-5.6,0-10.1-4.6-10.1-10.1
        c0-5.6,4.5-10.1,10.1-10.1
        C84.2,46.7,88.7,51.3,88.7,56.8z
      };
  }
}

\usepackage[
  colorlinks=true,
  citecolor=red,
  linkcolor=blue,
  urlcolor=blue
]{hyperref}
\newcommand{\orcid}[1]{%
  \href{https://orcid.org/#1}{%
    \mbox{%
      \scalerel*{%
        \begin{tikzpicture}[yscale=-1,transform shape]
          \pic{orcidlogo};
        \end{tikzpicture}%
      }{|}%
    }%
  }%
}

\newcommand{\beq}{\begin{equation}}
\newcommand{\eeq}{\end{equation}}
\newcommand{\bea}{\begin{eqnarray}}
\newcommand{\eea}{\end{eqnarray}}

\newcommand{\Boxop}{\Box}
\newcommand{\TT}{\mathrm{TT}}
\newcommand{\Lie}{\mathcal{L}}

\newcommand{\dd}{\mathrm{d}}

\newcommand{\comm}[1]{}

\begin{document}
\title{Sourced Gravitational Radiation in Parity-Violating Tetrad Gravity
}
\author{Gamal G. L. Nashed}\email{nashed@bue.edu.eg}
\affiliation{Centre for Theoretical Physics, The British University, P.O. Box 43, El Sherouk City, Cairo 11837, Egypt\\Centre for Space Research, North-West University, Potchefstroom 
2520, South Africa}

\date{\today}

\begin{abstract}
We study linear gravitational modes, sourced far-zone fields, and quadratic energy balance in a four-parameter tetrad theory of gravity. The theory contains three parity-even torsion invariants and a parity-odd coupling $v^\mu a_\mu$, where $v_\mu$ and $a_\mu$ are the vector and axial torsion components. Linearizing around Minkowski spacetime and applying a scalar--vector--tensor decomposition, we find two tensor, two scalar, and four vector degrees of freedom in the generic nondegenerate parity-violating region. The tensor and scalar modes propagate at the speed of light, while the two vector branches have coupling-dependent propagation speeds. We derive their kinetic eigenvalues, characteristic speeds, and flux normalizations. An exact trace identity shows that the vector Hamiltonian is not positive definite in the generic branch, even when the kinetic eigenvalues are positive and the propagation speeds are real. We also derive the response to a general compact canonical source and obtain explicit source-to-amplitude relations for all eight linear polarizations. The field $V_\mu=\partial^\nu B_{\nu\mu}$ is used to illustrate intrinsic-spin multipoles. {We further prove that the quadratic Noether current equals the second-order M{\o}ller current for the chosen first-derivative tetrad Lagrangian and derive the corresponding mode-resolved energy and momentum fluxes.} The vector contribution remains a formal signed response because of the unbounded Hamiltonian. The stationary limit reproduces the results of Ref.~\cite{ShirafujiNashed1997}, and a representative parameter scan illustrates vector-mode splitting and kinetic degeneracy.
\end{abstract}

\maketitle

\begin{center}
\textit{This work is dedicated to the memory of my late supervisor, Professor Takashi Shirafuji of Saitama University. We began exploring this problem before my graduation, but I was unable to complete it at that time. I now present this work in fulfillment of the scientific endeavor we once started together.}
\end{center}

\section{Introduction}

Gravitational-wave observations provide a direct way to test gravity in its dynamical regime. Extensions of general relativity may change the number of gravitational-wave polarizations, their propagation speeds, and their energy fluxes. Tetrad theories are well suited to study such effects because torsion can be decomposed into vector, axial, and tensor components. This allows both parity-even and parity-odd quadratic torsion invariants \cite{HayashiShirafuji1979,BlixtEtAl2019}.

The energy and momentum of isolated systems in the general four-parameter tetrad theory were studied in Ref.~\cite{ShirafujiNashed1997}. The theory contains three parity-even torsion invariants and one parity-violating coupling. For stationary asymptotically flat fields, the linearized equations were solved to order $1/r$, and M{\o}ller's superpotential gave
\begin{equation}
E=m,
\qquad
P_\alpha=\mathring B_\alpha.
\end{equation}
Here, $E$ and $P_\alpha$ are the total energy and spatial momentum, $m$ is the gravitational mass, and $\mathring B_\alpha$ is the constant vector in the stationary far-field solution.

The corresponding time-dependent problem requires a separate treatment. Time dependence introduces retarded fields, changes the asymptotic equations from elliptic to hyperbolic form, and produces energy and momentum fluxes at quadratic order. A complete scalar--vector--tensor decomposition is needed to separate constraints from propagating degrees of freedom.

Linear perturbations in new general relativity have been studied extensively. Reference~\cite{GolovnevEtAl2024} classified the scalar, pseudoscalar, vector, pseudovector, and tensor modes around the Minkowski tetrad in the parity-even theory. Reference~\cite{BahamondeEtAl2025} presented a gauge-invariant SVT and Hessian analysis, including the ghost-free parameter branches. The parity-violating extension has also been studied on FLRW backgrounds \cite{KangLiTong2025}. In addition, gravitational-wave energy and momentum in extended NGR were examined for weak Newtonian sources in Ref.~\cite{SakaneKawai2002}.

The purpose of this work is to study the generic time-dependent and asymptotically flat problem in the conventions of Ref.~\cite{ShirafujiNashed1997}. We derive a complete linear source-to-mode map for a compact canonical energy--momentum source. This includes the scalar constraint reduction, tensor and scalar retarded solutions, and vector responses obtained using a Schur complement and pole-residue projection. We also use the composite field
\begin{equation}
V_\mu=\partial^\nu B_{\nu\mu},
\end{equation}
to describe intrinsic-spin multipoles. Finally, we prove the M{\o}ller--Noether identity and derive the mode-resolved energy and momentum fluxes. The far-zone amplitudes are expressed in terms of source projections; explicit waveforms require a specified matter model.

Table~\ref{tab:comparison} compares the present work with related studies.

\begin{table}[H]
\centering
\caption{Comparison of related studies. PV denotes the parity-violating
torsion coupling.}
\label{tab:comparison}
\renewcommand{\arraystretch}{1.35}
\setlength{\tabcolsep}{9pt}
\resizebox{\textwidth}{!}{%
\begin{tabular}{llll}
\hline
\textbf{Work} &
\textbf{Background} &
\textbf{Theory} &
\textbf{Method}
\tabularnewline
\hline
Shirafuji--Nashed (1997) \cite{ShirafujiNashed1997} &
Stationary, asymptotically flat &
Four-parameter theory with PV &
Linear far-field analysis and M{\o}ller integral
\tabularnewline
\hline
Sakane--Kawai (2002) \cite{SakaneKawai2002} &
Weak-field radiation &
Extended NGR &
Wave-zone analysis
\tabularnewline
\hline
Bahamonde et al. (2025) \cite{BahamondeEtAl2025} &
Minkowski perturbations &
Parity-even NGR &
SVT and Hessian analysis
\tabularnewline
\hline
Kang--Li--Tong (2025) \cite{KangLiTong2025} &
FLRW perturbations &
Extended NGR with PV &
Helicity and SVT analysis
\tabularnewline
\hline
Present work &
Minkowski perturbations and wave zone &
Four-parameter theory with PV &
Sourced SVT, characteristics, and mode-resolved analysis
\tabularnewline
\hline
\end{tabular}%
}
\end{table}
The paper is organized as follows. Section~\ref{2} introduces the tetrad theory and derives the linearized field equations. Section~\ref{3} presents the SVT decomposition. Sections~\ref{4}--\ref{6} discuss the parameter mapping, vector characteristics, and special limits. Section~\ref{7} presents a representative scan of the reduced quadratic matrices. Section~\ref{sec:full-sourced-svt} derives the complete sourced SVT response. Sections~\ref{8}--\ref{10} study outgoing modes, intrinsic-spin multipoles, energy--momentum balance, and the stationary limit. Finally, Sec.~\ref{11} presents the discussion and conclusions.

\section{Tetrad Theory and Linearized Equations}
\label{2}

We use Greek indices $\mu,\nu,\ldots=0,1,2,3$ for spacetime coordinates, internal Lorentz indices $k,l,\ldots=0,1,2,3$, and Latin indices $i,j,\ldots=1,2,3$ for spatial coordinates. The Minkowski metric is
\begin{equation}
\eta_{\mu\nu}
=
\operatorname{diag}(-1,1,1,1),
\end{equation}
and $\epsilon_{0123}=+1$. Indices on perturbations are raised and lowered using $\eta_{\mu\nu}$. We write $X_{,\mu}=\partial_\mu X$, $\dot X=\partial_tX$,
\begin{equation}
\Delta
=
\delta^{ij}\partial_i\partial_j,
\qquad
\Boxop
=
\eta^{\mu\nu}\partial_\mu\partial_\nu.
\end{equation}

The tetrad $b^k{}_\mu$ defines the metric and Weitzenb\"ock connection:
\begin{equation}
g_{\mu\nu}
=
b_{k\mu}b^k{}_\nu,
\qquad
\Gamma^\lambda{}_{\mu\nu}
=
b_k{}^\lambda\partial_\nu b^k{}_\mu.
\end{equation}
The torsion tensor is
\begin{equation}
T^\lambda{}_{\mu\nu}
=
b_k{}^\lambda
\left(
\partial_\nu b^k{}_\mu
-
\partial_\mu b^k{}_\nu
\right).
\end{equation}
Its vector and axial components are
\begin{equation}
v_\mu
=
T^\lambda{}_{\lambda\mu},
\qquad
a_\mu
=
\frac16
\epsilon_{\mu\nu\rho\sigma}
T^{\nu\rho\sigma}.
\end{equation}
\begingroup
The remaining irreducible tensor component is
\begin{equation}
t_{\lambda\mu\nu}
=
\frac12\left(T_{\lambda\mu\nu}+T_{\mu\lambda\nu}\right)
+\frac16\left(g_{\nu\lambda}v_\mu+g_{\nu\mu}v_\lambda\right)
-\frac13g_{\lambda\mu}v_\nu .
\label{eq:tensor-torsion-definition}
\end{equation}
Thus every torsion quantity appearing in the action below has now been
defined explicitly.
\endgroup

The four-parameter gravitational Lagrangian is
\begin{equation}
\Lie_G
=
\frac{\sqrt{-g}}{\kappa}
\left[
a_1t^{\mu\nu\lambda}t_{\mu\nu\lambda}
+
a_2v^\mu v_\mu
+
a_3a^\mu a_\mu
+
a_4v^\mu a_\mu
\right],
\label{eq:lagrangian}
\end{equation}
where $\kappa=8\pi G$. The first three terms are parity even, while the term proportional to $a_4$ violates parity.

The Newtonian limit requires
\begin{equation}
a_1+4a_2+9a_1a_2=0.
\label{eq:newton}
\end{equation}
We parameterize the remaining parity-even freedom by
\begin{equation}
a_1
=
-\frac{1}{3(1-\epsilon)},
\qquad
a_2
=
\frac{1}{3(1-4\epsilon)}.
\label{eq:epsilon}
\end{equation}
The values $\epsilon=1/4$ and $\epsilon=1$ are excluded because the coefficients become singular.

Around Minkowski spacetime, we write
\begin{equation}
b^k{}_\mu
=
\delta^k{}_\mu+a^k{}_\mu,
\qquad
a_{\mu\nu}
=
\frac12h_{\mu\nu}+A_{\mu\nu}, \quad \mbox{where} \quad
h_{\mu\nu}=h_{\nu\mu},
\qquad
A_{\mu\nu}=-A_{\nu\mu}.
\end{equation}
The metric perturbation is
\begin{equation}
g_{\mu\nu}
=
\eta_{\mu\nu}+h_{\mu\nu}.
\end{equation}
The rank-two tetrad perturbation $a_{\mu\nu}$ should not be confused with the axial torsion vector $a_\mu$.

We define the trace-reversed metric perturbation and dual antisymmetric field by
\begin{equation}
\bar h_{\mu\nu}
=
h_{\mu\nu}
-\frac12\eta_{\mu\nu}h,
\qquad
\bar A_{\mu\nu}
=
\frac12
\epsilon_{\mu\nu\rho\sigma}
A^{\rho\sigma},\quad \mbox{and impose the harmonic gauge condition} \quad
\partial_\nu\bar h^{\mu\nu}=0.
\end{equation}

To first order, the torsion tensor is
\begin{equation}
T^\lambda{}_{\mu\nu}
=
\frac12
\left(
\partial_\nu h^\lambda{}_\mu
-
\partial_\mu h^\lambda{}_\nu
\right)
+
\partial_\nu A^\lambda{}_\mu
-
\partial_\mu A^\lambda{}_\nu.
\label{eq:linearTsplit}
\end{equation}
The torsion vector becomes
\begin{equation}
v_\mu
=
\frac12
\left(
\partial_\mu h
-
\partial^\lambda h_{\lambda\mu}
\right)
-
\partial^\lambda A_{\lambda\mu}, \quad \mbox{or, in harmonic gauge,} \quad
v_\mu
=
\frac14\partial_\mu h
-
\partial^\lambda A_{\lambda\mu}.
\label{eq:linearvharmonic}
\end{equation}
Therefore, the symmetric metric perturbation and antisymmetric tetrad perturbation remain coupled even around Minkowski spacetime.

\subsection{Quadratic Action and Linear Field Equations}

Since the Minkowski background has zero torsion, the torsion components begin at first order. The quadratic action is therefore
\begin{equation}
S_G^{(2)}
=
\frac1\kappa
\int\dd^4x
\left[
a_1t^{(1)\mu\nu\lambda}t^{(1)}_{\mu\nu\lambda}
+
a_2v^{(1)\mu}v^{(1)}_\mu
+
a_3a^{(1)\mu}a^{(1)}_\mu
+
a_4v^{(1)\mu}a^{(1)}_\mu
\right].
\label{eq:quadraticaction}
\end{equation}

Varying this action gives
\begin{equation}
\partial_\nu
\mathcal P_\lambda{}^{\mu\nu}
=
0,\quad \mbox{where} \quad  \mathcal P_\lambda{}^{\mu\nu}
=
\frac{\partial\mathcal L_G^{(2)}}
{\partial T^{(1)\lambda}{}_{\mu\nu}},
\label{eq:linearEuler}
\end{equation}
is antisymmetric in $\mu$ and $\nu$. The symmetric and antisymmetric parts of this equation describe the equations for $h_{\mu\nu}$ and $A_{\mu\nu}$, respectively.

In harmonic gauge, the vacuum field equations are
\begin{align}
\frac{3a_1}{2}\Boxop\bar h_{\mu\nu}
&-
\frac{a_1+a_2}{2}
\left(
\eta_{\mu\nu}\Boxop\bar h
-\partial_\mu\partial_\nu\bar h
\right)
+
2(a_1+a_2)
{A^\rho}_{(\mu,\nu)\rho}
+
\frac{2a_4}{3}
{\bar A^\rho}_{(\mu,\nu)\rho}
=
0,
\label{eq:symmetric}
\\
\left(
a_1-\frac{4a_3}{9}
\right)
\Boxop A_{\mu\nu}
&-
2\left(
a_2+\frac{4a_3}{9}
\right)
{A^\rho}_{[\mu,\nu]\rho}
-
\frac{2a_4}{3}
{\bar A^\rho}_{[\mu,\nu]\rho}
-
\frac{a_4}{3}
\epsilon_{\mu\nu\lambda\rho}
\partial^\lambda\partial_\sigma
A^{\rho\sigma}
=
0.
\label{eq:antisymmetric}
\end{align}

Define
\begin{equation}
s=a_1+a_2,
\qquad
\widetilde a_4=\frac{a_4}{3},
\qquad
B_{\mu\nu}
=
sA_{\mu\nu}
+
\widetilde a_4\bar A_{\mu\nu}.
\label{eq:Bdefinition}
\end{equation}
For $s^2+\widetilde a_4^2\neq0$,
\begin{equation}
A_{\mu\nu}
=
\frac{
sB_{\mu\nu}
-
\widetilde a_4\bar B_{\mu\nu}
}{
s^2+\widetilde a_4^2
}.
\end{equation}
Here, the bar denotes the Hodge dual of an antisymmetric tensor. The tensor $B_{\mu\nu}$ is distinct from the transverse pseudovector $B_i$ introduced in the later SVT decomposition.

The divergence of the coupled antisymmetric field,
\begin{equation}
V_\mu
=
\partial^\nu B_{\nu\mu},\quad \mbox{satisfies} \quad
\Boxop V_\mu=0,
\qquad
\partial^\mu V_\mu=0.
\label{eq:Vwave}
\end{equation}
Thus, $V_\mu$ behaves as a transverse massless field.

With matter, the corresponding equation is
\begin{equation}
\Boxop V_\mu
=
\kappa\partial^\nu T_{[\mu\nu]}.
\label{eq:Vsourced}
\end{equation}
For a source with intrinsic spin,
\begin{equation}
T_{[\mu\nu]}
=
\frac12
\partial_\lambda S_{\mu\nu}{}^\lambda.
\end{equation}
Therefore, the field $V_\mu$ is sourced by derivatives of the intrinsic-spin current. Equation~\eqref{eq:Vsourced} applies only to the composite field $V_\mu$; it does not directly determine every vector eigenmode. The nonluminal vector modes must instead be obtained by projecting the full tetrad source onto the eigenvectors of the reduced vector system.

\section{SVT Decomposition}
\label{3}

We decompose the antisymmetric tetrad perturbation into scalar, pseudoscalar, vector, and pseudovector parts:
\begin{equation}
A_{0i}
=
C_i+\partial_i\mu,
\qquad
A_{ij}
=
\epsilon_{ijk}
\left(
B_k+\partial_k\chi
\right),
\qquad
\partial_iC_i=\partial_iB_i=0.
\label{eq:SVTantisymmetric}
\end{equation}
Here, $\mu$ is a scalar, $\chi$ is a pseudoscalar, $C_i$ is a transverse vector, and $B_i$ is a transverse pseudovector.

The symmetric metric perturbation is decomposed as
\begin{align}
h_{00}
&=
2\phi,
\nonumber\\
h_{0i}
&=
\partial_i\mathcal B+S_i,
\nonumber\\
h_{ij}
&=
2\psi\delta_{ij}
+
2\partial_i\partial_j\mathcal E
+
\partial_iF_j+\partial_jF_i
+
h_{ij}^{\TT},
\label{eq:SVTsymmetric}
\end{align}
where
\begin{equation}
\partial_iS_i
=
\partial_iF_i
=
0,
\qquad
\partial_ih_{ij}^{\TT}
=
0,
\qquad
h_{ii}^{\TT}
=
0.
\end{equation}
The fields $\phi$, $\mathcal B$, $\psi$, and $\mathcal E$ are metric scalars; $S_i$ and $F_i$ are transverse vectors; and $h_{ij}^{\TT}$ is the transverse-traceless tensor perturbation. We use $\mathcal E$ instead of $E$ to avoid confusion with the total energy.

For nonzero spatial momentum, this decomposition is unique. Under an infinitesimal coordinate transformation,
\begin{equation}
\xi^0=T,
\qquad
\xi^i=\partial_iL+L_i,
\qquad
\partial_iL_i=0,
\end{equation}
the metric variables transform, but gauge-invariant combinations can be constructed. The antisymmetric variables remain independent because fixing the inertial spin connection in the Weitzenb\"ock gauge does not introduce six additional tetrad gauge transformations.

Substituting the SVT decomposition into the linear torsion tensor gives
\begin{align}
v_0
&=
3\dot\psi
+
\Delta
\left(
\dot{\mathcal E}
-\frac12\mathcal B+\mu
\right),
\qquad
v_i^{(S)}
=
\partial_i
\left(
-\phi+2\psi+\frac12\dot{\mathcal B}+\dot\mu
\right),
\nonumber\\
v_i^{(V)}
&=
\frac12
\left(
\dot S_i-\Delta F_i
\right)
+
\dot C_i
+
(\bm\nabla\times\bm B)_i,
\end{align}
and
\begin{align}
a_0
=
-\frac23\Delta\chi,
\quad
a_i^{(S)}
=
-\frac23\partial_i\dot\chi,
\quad
a_i^{(V)}
=
-\frac23
\left[
\dot B_i+(\bm\nabla\times\bm C)_i
\right].
\end{align}

The parity-even part of the action does not mix scalar and vector sectors. The parity-odd term $a_4v^\mu a_\mu$ mixes scalar and pseudoscalar variables, and also mixes vector and pseudovector variables. It does not contribute to the tensor action on Minkowski spacetime.

\subsection{Scalar Constraint Reduction}

For the scalar sector, we use the two scalar gauge functions to impose
\begin{equation}
\mathcal E=0,
\qquad
\mathcal B=0.
\label{eq:scalargauge}
\end{equation}
We use the Fourier convention
\begin{equation}
X(t,\bm x)
=
\int
\frac{\dd^3k}{(2\pi)^3}
X(t,\bm k)e^{i\bm k\cdot\bm x},
\end{equation}
with $k=|\bm k|>0$. The remaining scalar variables are $\phi$, $\psi$, $\mu$, and $\chi$.

\begingroup
Direct substitution into Eq.~\eqref{eq:quadraticaction} gives, apart
from the common Fourier measure,
\begin{align}
\mathcal L_S={}&\frac1\kappa\Bigg[
-9a_2\dot\psi^2+s k^2\dot\mu^2
-k^2\left(a_1-\frac{4a_3}{9}\right)\dot\chi^2
-\frac{2a_4}{3}k^2\dot\chi\dot\mu
+s k^2\phi^2
+2(a_1-2a_2)k^2\phi\psi
-2s k^2\phi\dot\mu
+\frac{2a_4}{3}k^2\phi\dot\chi
\nonumber\\
&+(a_1+4a_2)k^2\psi^2
-2(a_1-2a_2)k^2\psi\dot\mu
-\frac{4a_4}{3}k^2\psi\dot\chi
+6a_2k^2\mu\dot\psi-2a_4k^2\chi\dot\psi
-s k^4\mu^2
+\frac{2a_4}{3}k^4\chi\mu
+\left(a_1-\frac{4a_3}{9}\right)k^4\chi^2
\Bigg].
\label{eq:scalarunreduced}
\end{align}
This unreduced expression makes the two successive constraints and all
signs in the reduced action independently checkable.
\endgroup

The lapse perturbation $\phi$ has no time derivative and is therefore a constraint. Its equation gives
\begin{equation}
\phi
=
\dot\mu
-
\frac{a_1-2a_2}{s}\psi
-
\frac{a_4}{3s}\dot\chi,
\qquad
s=a_1+a_2.
\label{eq:phiconstraint}
\end{equation}
After substituting this relation, $\mu$ also becomes nondynamical:
\begin{equation}
\mu
=
\frac{
9a_2\dot\psi+a_4k^2\chi
}{
3sk^2
}.
\label{eq:muconstraint}
\end{equation}

The reduced scalar action is
\begin{equation}
S_S^{(2)}
=
-\frac1\kappa
\int\dd t\,\dd^3k
\left[
\frac{9a_1a_2}{s}
\left(
\dot\psi^2-k^2\psi^2
\right)
+
\frac{f_1}{s}k^2
\left(
\dot\chi^2-k^2\chi^2
\right)
\right], \quad \mbox{where} \quad f_1
=
s
\left(
a_1-\frac{4a_3}{9}
\right)
+
\frac{a_4^2}{9}.
\label{eq:scalarreduced}
\end{equation}
\begingroup
The symbolic reduction agrees with Eq.~\eqref{eq:scalarreduced} up to
the boundary term
\begin{equation}
-\frac{2a_1a_4k^2}{s}\frac{\dd}{\dd t}(\psi\chi),
\label{eq:scalar-boundary-term}
\end{equation}
which does not modify the Euler--Lagrange equations. This term is
retained by the computer-algebra comparison before being removed by
integration by parts.
\endgroup
The canonically normalized pseudoscalar variable is $\widehat\chi=k\chi$.

The scalar field equations are
\begin{equation}
\frac{a_1a_2}{s}\Boxop\psi=0,
\qquad
\frac{f_1}{s}\Boxop\chi=0.
\label{eq:scalarwaves}
\end{equation}
Both scalar modes therefore propagate at the speed of light. The no-ghost conditions are
\begin{equation}
-\frac{a_1a_2}{s}>0,
\qquad
-\frac{f_1}{s}>0.
\label{eq:scalarghostconditions}
\end{equation}

\subsection{Mode Counting and Degenerate Cases}

For generic nondegenerate parameters, the theory contains the following physical modes:

\begin{table}[H]
\centering
\caption{Generic linear degrees of freedom around Minkowski spacetime.}
\label{tab:svtcount}
\renewcommand{\arraystretch}{1.3}
\setlength{\tabcolsep}{8pt}
\begin{tabular}{lccc}
\hline
\textbf{Sector}
&
\textbf{Initial variables}
&
\textbf{Physical fields}
&
\textbf{Polarizations}
\tabularnewline
\hline
Tensor
&
\(h_{ij}^{\TT}\)
&
2
&
\(+,\times\)
\tabularnewline
\hline
Scalar
&
\(\phi,\mathcal B,\psi,\mathcal E,\mu,\chi\)
&
2
&
Scalar and pseudoscalar
\tabularnewline
\hline
Vector
&
\(S_i,F_i,C_i,B_i\)
&
4
&
Two transverse branches
\tabularnewline
\hline
\end{tabular}
\end{table}
The reduction follows three steps:
\begin{enumerate}
\item Fix the coordinate gauge using the four diffeomorphism functions.
\item Solve the equations for nondynamical variables.
\item Diagonalize the remaining kinetic and gradient matrices.
\end{enumerate}

The generic mode count does not apply when the constraint or kinetic structure becomes degenerate. The important degenerate surfaces are
\begin{equation}
s=0,
\qquad
c=a_1-\frac{4a_3}{9}=0,
\qquad
f_1=0,
\qquad
\mathcal A_V=0.
\label{eq:degeneracysurfaces}
\end{equation}
On these surfaces, the rank of the constraint system or kinetic matrix changes, and a separate analysis is required.

\section{Relation to the Modern Invariant Basis}
\label{4}

To compare our formulation with recent stability analyses, we rewrite the action using the standard quadratic torsion invariants with coefficients $c_i$ \cite{BahamondeEtAl2025,KangLiTong2025}. With the conventions of Ref.~\cite{KangLiTong2025}, the parameters are related by
\begin{equation}
c_1
=
-a_1+\frac{a_3}{9},
\qquad
c_2
=
-a_1-\frac{2a_3}{9},
\qquad
c_3
=
a_1-2a_2,
\qquad
c_4
=
-\frac{2a_4}{3}.
\label{eq:mapping}
\end{equation}

This relation follows from expressing the standard torsion contractions,
\begin{equation}
I_1
=
T_{\rho\mu\nu}T^{\rho\mu\nu},
\qquad
I_2
=
T_{\rho\mu\nu}T^{\nu\mu\rho},
\qquad
I_3
=
T_\rho T^\rho,
\end{equation}
in terms of the irreducible tensor, vector, and axial torsion components. Here,
\begin{equation}
T_\rho
=
T^\lambda{}_{\lambda\rho},\quad \mbox{and the parity-odd invariant is defined by} \quad
I_{\rm odd}
=
-\frac12
\epsilon^{\mu\nu\rho\sigma}
T_\mu T_{\nu\rho\sigma}.
\label{eq:Iodddefinition}
\end{equation}

Matching the coefficients of $t^2$, $v^2$, $a^2$, and $v^\mu a_\mu$ gives Eq.~\eqref{eq:mapping}. Two useful consistency relations are
\begin{equation}
2c_1+c_2
=
-3a_1,
\qquad
Z
\equiv
2c_1+c_2+c_3
=
-2(a_1+a_2).
\end{equation}

\begingroup
The tensor and generic scalar kinetic terms, together with a nonempty
vector kinetic-positive region, require
\begin{equation}
a_1<0,
\qquad
s=a_1+a_2>0,
\qquad
c
=
a_1-\frac{4a_3}{9}
<0.
\label{eq:signregion}
\end{equation}

In the generic nondegenerate parity-violating branch, the parity-odd coupling must also satisfy
\begin{equation}
a_4^2
<
\min
\left\{
M_1^2,M_2^2,M_3^2
\right\}, \quad \mbox{where} \quad M_1^2
=
-9sc, \quad M_2^2
=
-\left(
9a_1^2
+
36a_1a_2
+
8a_1a_3
-
4a_2a_3
\right),\quad \mbox{and}  \quad M_3^2
=
-\frac{27a_1sc}
{7a_1-\dfrac{16a_3}{9}}.
\label{eq:a4bound}
\end{equation}
The quantities $M_1^2$, $M_2^2$, and $M_3^2$ are coupling combinations, not physical masses. In the sign region of Eq.~\eqref{eq:signregion}, a nonempty interval additionally requires these upper bounds to be positive. They determine the parameter interval for which the pseudoscalar and vector kinetic terms have the required signs. The equality
\begin{equation}
a_4^2=M_3^2, \quad \mbox{defines a kinetic-degeneracy boundary.}
\end{equation}

These kinetic conditions are necessary but not sufficient for complete stability. As shown in Sec.~\ref{sec:vector-boundedness}, the gradient contribution must also give a positive Hamiltonian. In the generic parity-violating branch, this additional requirement is not satisfied.
\endgroup

\section{Vector Characteristics}
\label{5}

\subsection{Vector Variables and Constraint}

We analyze the vector sector in the standard $c_i$ basis defined in Eq.~\eqref{eq:mapping}. We use the transverse coordinate freedom to set
\begin{equation}
F_i=0.
\end{equation}
For each transverse helicity mode, we define
\begin{equation}
\beta_i
=
-\frac12S_i-C_i,
\qquad
\gamma_i
=
\frac12S_i-C_i,
\qquad
\lambda_i
=
B_i.
\label{eq:vectorvariablemap}
\end{equation}
We expand the transverse vectors into left- and right-handed circular polarizations, labelled by $A=L,R$, with
\begin{equation}
p_L=-1,
\qquad
p_R=+1.
\end{equation}

The vector action contains three amplitudes, $\beta_A$, $\gamma_A$, and $\lambda_A$.
\begingroup
Before eliminating the nondynamical amplitude, the action for each
helicity is
\begin{align}
S_V^{(2)}
=&-\frac1{2\kappa}\sum_A\int\dd t\,\dd^3k\,
\Big\{
Z\dot\beta_A^2+Zk^2\lambda_A^2
-2c_1k^2\beta_A^2+2c_2k^2\beta_A\gamma_A
-2c_1k^2\gamma_A^2+(2c_2-4c_1)\dot\lambda_A^2
+(2c_2-4c_1)p_Ak\gamma_A\dot\lambda_A
\nonumber\\
&
+(4c_2+2c_3)p_Ak\beta_A\dot\lambda_A
+c_4\left(
2\dot\beta_A\dot\lambda_A-k^2\beta_A\lambda_A
-k^2\lambda_A\gamma_A-p_Ak\gamma_A\dot\beta_A
\right)\Big\},
\qquad Z=2c_1+c_2+c_3.
\label{eq:vectorunreduced}
\end{align}
The absence of \(\dot\gamma_A\) is manifest.
\endgroup
Thus, $\gamma_A$ is nondynamical, and its constraint equation is
\begin{equation}
-4c_1k^2\gamma_A
+
2c_2k^2\beta_A
+
2(c_2-2c_1)p_Ak\dot\lambda_A
-
c_4k^2\lambda_A
-
c_4p_Ak\dot\beta_A
=
0.
\label{eq:gammaconstraint}
\end{equation}
For $c_1\neq0$, this gives
\begin{equation}
\gamma_A
=
\frac{
2c_2k^2\beta_A
+
2(c_2-2c_1)p_Ak\dot\lambda_A
-
c_4k^2\lambda_A
-
c_4p_Ak\dot\beta_A
}{
4c_1k^2
}.
\label{eq:gammasolution}
\end{equation}

After eliminating $\gamma_A$, the two propagating vector variables are
\begin{equation}
q_A
=
\begin{pmatrix}
\beta_A\\
\lambda_A
\end{pmatrix}.
\end{equation}
Their kinetic action is
\begin{equation}
S_{V,\mathrm{kin}}^{(2)}
=
\frac1{2\kappa}
\sum_A
\int\dd t\,\dd^3k\,
\dot q_A^T\bm M_V\dot q_A,
\end{equation}
where
\begin{equation}
\bm M_V
=
-\frac1{8c_1}
\begin{pmatrix}
8c_1Z+c_4^2 &
2c_4(6c_1-c_2)
\\
2c_4(6c_1-c_2) &
4(c_2^2-4c_1^2)
\end{pmatrix},
\qquad
Z=2c_1+c_2+c_3.
\label{eq:vectorkineticmatrix}
\end{equation}

\subsection{Kinetic Conditions}

In terms of the original parameters, the kinetic matrix is
\begin{equation}
\bm M_V
=
\begin{pmatrix}
\dfrac{
36a_1^2+36a_1a_2-4a_1a_3-4a_2a_3+a_4^2
}{
2(9a_1-a_3)
}
&
\dfrac{
a_4(45a_1-8a_3)
}{
6(9a_1-a_3)
}
\\[4mm]
\dfrac{
a_4(45a_1-8a_3)
}{
6(9a_1-a_3)
}
&
-\dfrac{
3a_1(9a_1-4a_3)
}{
2(9a_1-a_3)
}
\end{pmatrix}.
\label{eq:MVoriginal}
\end{equation}
The vector sector has positive kinetic energy only if
\begin{equation}
\operatorname{tr}\bm M_V>0,
\qquad
\det\bm M_V>0.
\label{eq:vectorghostconditions}
\end{equation}
The determinant vanishes when
\begin{equation}
a_4^2=M_3^2,
\end{equation}
which defines a kinetic-degeneracy surface. On this surface, one vector combination loses its quadratic kinetic term, and the generic vector analysis is no longer valid.

\subsection{Propagation Speeds}

The two vector amplitudes satisfy
\begin{equation}
\mathcal K_{V,A}(\omega,k)q_A=0.
\end{equation}
Both helicities have the same characteristic equation:
\begingroup
\begin{equation}
27a_1cs(k^2-\omega^2)^2
+a_4^2\left[
\left(7a_1-\frac{16a_3}{9}\right)\omega^4
-(10a_1+4a_2)k^2\omega^2
+3a_1k^4
\right]
=
0,\quad \mbox{where} \quad s=a_1+a_2,
\quad
c=a_1-\frac{4a_3}{9}.
\label{eq:vectorcharacteristic}
\end{equation}
\endgroup
Defining $x=\omega^2/k^2$, the characteristic equation becomes
\begingroup
\begin{align}
\mathcal A_Vx^2
+
\mathcal B_Vx
+
\mathcal C_V
=
0, \quad \mbox{with} \quad
\mathcal A_V
=
27a_1cs+a_4^2\left(7a_1-\frac{16a_3}{9}\right),
\quad
\mathcal B_V
=
-54a_1cs-(10a_1+4a_2)a_4^2,
\quad
\mathcal C_V
=
27a_1cs+3a_1a_4^2.
\label{eq:vectorABC}
\end{align}
\endgroup
The two squared propagation speeds are
\begin{equation}
c_{V\pm}^2
=
\frac{
-\mathcal B_V
\pm
\sqrt{
\mathcal B_V^2
-
4\mathcal A_V\mathcal C_V
}
}{
2\mathcal A_V
}.
\label{eq:vectorspeeds}
\end{equation}
Real positive values of $c_{V\pm}^2$ are required for hyperbolic propagation. The parity-odd coupling $a_4$ splits the two vector branches, but the two helicities of a given branch have the same propagation speed on Minkowski spacetime.

\begingroup
As a reproducibility check, the accompanying symbolic-algebra notebook
encodes the scalar and vector component contractions obtained by substituting
the SVT tetrad into Eq.~\eqref{eq:quadraticaction}, and performs the constraint
eliminations as Schur complements. It verifies
Eq.~\eqref{eq:scalarreduced}, modulo the boundary term
\eqref{eq:scalar-boundary-term}, and returns an identically vanishing
matrix residual for Eq.~\eqref{eq:vectorkineticmatrix}. Finally, it evaluates
\begin{equation}
\det\!\left[
\bm G_V-x\bm M_V+2ip_A\sqrt{x}\,\bm J_V
\right]
\end{equation}
and reproduces Eqs.~\eqref{eq:vectorcharacteristic}--\eqref{eq:vectorspeeds}.
All checks are symbolic and are made before assigning numerical values to
the couplings.
\endgroup

\begingroup
The symmetric gradient matrix obtained from the same constraint
reduction is
\begin{equation}
\bm G_V=\frac1{8c_1}
\begin{pmatrix}
4(c_2^2-4c_1^2)&-2(2c_1+c_2)c_4\\
-2(2c_1+c_2)c_4&8c_1Z+c_4^2
\end{pmatrix}.
\label{eq:vectorG}
\end{equation}
\endgroup

\subsection{Hamiltonian Boundedness}
\label{sec:vector-boundedness}

Real propagation speeds and positive kinetic eigenvalues are not sufficient to ensure stability. The vector Hamiltonian is bounded from below only if
\begin{equation}
\bm M_V>0,
\qquad
\bm G_V>0,
\label{eq:fullvectorpositivity}
\end{equation}
where $\bm G_V$ is the gradient matrix. The kinetic and gradient matrices satisfy the exact identity
\begin{equation}
\operatorname{tr}\bm G_V
=
-\operatorname{tr}\bm M_V.
\label{eq:trace-nogo}
\end{equation}
Therefore, both matrices cannot be positive definite at the same time. If $\bm M_V$ is positive definite, then $\operatorname{tr}\bm G_V<0$, so $\bm G_V$ has at least one negative eigenvalue. Conversely, if $\bm G_V$ is positive definite, then $\bm M_V$ has a negative direction.

Thus, the generic vector Hamiltonian is unbounded from below for all values of the four couplings. The parity-odd coupling modifies the vector speeds and mixing, but it does not create an open stable vector region. Stable parity-even theories occur only on special degenerate branches where the vector modes disappear.

\subsection{Mode Normalization and Flux}

The reduced vector action can be written as
\begin{equation}
S_{V,A}^{(2)}
=
\frac1{2\kappa}
\int\dd t\,\dd^3k
\left[
\dot q_A^\dagger\bm M_V\dot q_A
+
2p_Ak\,q_A^\dagger\bm J_V\dot q_A
-
k^2q_A^\dagger\bm G_Vq_A
\right].
\label{eq:vectorMGJ}
\end{equation}
\begingroup
The antisymmetric gyroscopic matrix is
\begin{equation}
\bm J_V=
\begin{pmatrix}
0&-\dfrac{D_V}{16c_1}\\[1mm]
\dfrac{D_V}{16c_1}&0
\end{pmatrix},
\qquad
D_V=8c_1(c_2+c_3)+4c_2^2-c_4^2.
\label{eq:vectorJ}
\end{equation}
Equations~\eqref{eq:vectorkineticmatrix} and \eqref{eq:vectorG}
directly give Eq.~\eqref{eq:trace-nogo}; hence the no-go statement is
an algebraic identity, rather than an inference from the numerical
scan.
\endgroup
The matrix $\bm J_V$ is antisymmetric and affects the propagation eigenvectors, but it does not contribute to the conserved quadratic energy:
\begin{equation}
E_{V,A}^{(2)}
=
\frac1{2\kappa}
\int\dd^3k
\left[
\dot q_A^\dagger\bm M_V\dot q_A
+
k^2q_A^\dagger\bm G_Vq_A
\right].
\label{eq:vectorenergy}
\end{equation}

For an outgoing vector mode, let
\begin{equation}
\bm D_A(c)
=
\bm G_V-c^2\bm M_V+2ip_Ac\,\bm J_V,\quad \mbox{and let $e_\sigma^{(A)}$ satisfy}\quad \bm D_A(c_{V\sigma})
e_\sigma^{(A)}
=
0.
\label{eq:vectorpropagationmatrix}
\end{equation}
The far-zone solution has the form
\begin{equation}
q_A
=
\frac{
e_\sigma^{(A)}
}{r}
F_{\sigma A}
\left(
t-\frac{r}{c_{V\sigma}},
\bm n
\right)
+
O(r^{-2}).
\end{equation}
Its signed normalization is
\begin{equation}
Z_{V\sigma}^{(A)}
=
e_\sigma^{(A)\dagger}
\left(
\bm M_V
-
\frac{ip_A}{c_{V\sigma}}\bm J_V
\right)
e_\sigma^{(A)}.
\label{eq:vectorZ}
\end{equation}
The corresponding energy flux is
\begin{equation}
\frac{\dd E_{V\sigma A}}
{\dd t\,\dd\Omega}
=
\frac1\kappa
Z_{V\sigma}^{(A)}
c_{V\sigma}
\left|
\dot F_{\sigma A}
\right|^2.
\label{eq:vectorfluxexplicit}
\end{equation}

A field rescaling can change the magnitude of $Z_{V\sigma}^{(A)}$, but not its sign. A negative value therefore identifies a negative-energy vector mode. Since the generic vector Hamiltonian is unbounded, these fluxes should be interpreted as formal signed responses rather than stable radiative states.

\section{Special Parameter Limits and Rank Changes}
\label{6}

The generic scalar and vector reductions assume that the coefficients used to solve the constraint equations are nonzero. On special parameter surfaces, the rank of the system changes, and the reduced formulas are no longer valid. These cases must be analyzed directly from the unreduced actions in Eqs.~\eqref{eq:scalarunreduced} and \eqref{eq:vectorunreduced}.

\subsection{Parity-Even Generic Theory}

For $a_4=0$, with $s\neq0$ and $c\neq0$, the scalar coefficient becomes
\begin{equation}
f_1=sc.
\end{equation}
The two scalar modes remain luminal, with no-ghost conditions
\begin{equation}
-\frac{a_1a_2}{s}>0,
\qquad
-c>0.
\label{eq:parityevenscalar}
\end{equation}

The vector kinetic matrix becomes diagonal:
\begin{equation}
\left.\bm M_V\right|_{a_4=0}
=
\begin{pmatrix}
2s & 0 \\
0 & -\dfrac{27a_1c}{2(9a_1-a_3)}
\end{pmatrix}.
\label{eq:parityevenMV}
\end{equation}
The two vector branches have the same luminal characteristic,
\begin{equation}
\omega^2=k^2.
\end{equation}
However, coincident propagation speeds do not remove the vector modes when both kinetic eigenvalues are nonzero.

Positive kinetic eigenvalues are not sufficient for stability. As shown by Eq.~\eqref{eq:trace-nogo}, the gradient matrix has negative directions whenever the kinetic matrix is positive definite. Therefore, the generic parity-even vector sector has an unbounded Hamiltonian, in agreement with Ref.~\cite{BahamondeEtAl2025}.

\subsection{TEGR Limit}

In the standard torsion basis, TEGR satisfies
\begin{equation}
c_1:c_2:c_3=1:2:-4,
\qquad
c_4=0.
\end{equation}
Using Eq.~\eqref{eq:mapping}, this gives
\begin{equation}
a_2=-a_1,
\qquad
a_3=\frac94a_1,
\qquad
a_4=0, \quad \mbox{or} \quad
s=0,
\qquad
c=0,
\qquad
a_4=0.
\label{eq:TEGRsurface}
\end{equation}
With the usual normalization,
\begin{equation}
a_1=-\frac13,
\qquad
a_2=\frac13,
\qquad
a_3=-\frac34,
\qquad
a_4=0.
\end{equation}

On this surface, the antisymmetric field equation vanishes identically. The antisymmetric tetrad perturbation becomes a Lorentz-gauge degree of freedom, not a propagating field. The generic scalar and vector formulas cannot be used because they contain divisions by $s$, $c$, or $f_1$. Only the two transverse-traceless tensor polarizations of general relativity remain.

\subsection{The Surface $s=0$}

Let
\begin{equation}
a_2=-a_1,
\qquad
a_4\neq0.
\end{equation}
The scalar variables $\phi$ and $\mu$ become Lagrange multipliers. Their constraint equations are
\begin{equation}
6a_1\psi+\frac{2a_4}{3}\dot\chi=0,
\qquad
-6a_1\dot\psi+\frac{2a_4}{3}k^2\chi=0.
\label{eq:szeroconstraints}
\end{equation}
These imply
\begin{equation}
\ddot\chi+k^2\chi=0.
\label{eq:szeroscalar}
\end{equation}
Therefore, this rank-changing surface contains one luminal scalar combination, rather than the two scalar modes of the generic $s\neq0$ theory.

If $s=0$ and $a_4=0$, but $c\neq0$, the antisymmetric tensor has the gauge symmetry
\begin{equation}
A_{\mu\nu}
\longrightarrow
A_{\mu\nu}
+
\partial_{[\mu}\zeta_{\nu]}.
\end{equation}
Its physical content is one massless pseudoscalar mode, dual to a four-dimensional two-form. Imposing $c=0$ in addition gives the TEGR limit and removes this mode.

\subsection{Scalar and Vector Degeneracies}

The scalar coefficient satisfies
\begin{equation}
f_1
=
sc+\frac{a_4^2}{9}
=
\frac{a_4^2-M_1^2}{9}.
\label{eq:f1M1}
\end{equation}
Therefore, the surface
\begin{equation}
a_4^2=M_1^2
\end{equation}
makes the pseudoscalar kinetic term vanish. The perturbative theory can then become strongly coupled, and a nonlinear analysis is required.

Similarly,
\begin{equation}
a_4^2=M_3^2
\end{equation}
gives
\begin{equation}
\det\bm M_V=0,
\end{equation}
so one vector combination loses its kinetic term. This is a rank-changing boundary and requires a separate reduced-rank analysis.

If
\begin{equation}
c=0,
\qquad
a_4\neq0, \quad \mbox{ the vector characteristic equation reduces to} \quad
3a_1a_4^2
(k^2+\omega^2)^2
=
0.
\end{equation}
For real spatial momentum, this gives $\omega^2=-k^2$, which is not a hyperbolic propagating mode. The case $c=0$ and $a_4=0$ must instead be analyzed from the unreduced equations.

\begingroup
In the parity-even theory this last limit contains two distinct
rank-reduced branches identified in
Ref.~\cite{BahamondeEtAl2025}.  For
\(a_4=0\), \(c=0\), and generic \(a_2\), the vector and
pseudoscalar sectors vanish at linear order, while the spectrum contains
the two tensor modes and one massless scalar, subject to the corresponding
kinetic sign.  Imposing \(a_2=0\) as well removes that scalar and leaves
only the two tensor modes.  These branches are distinct from TEGR:
TEGR instead satisfies \(s=0=c\).  Therefore the statement that stable
parity-even branches have no propagating vectors does not imply that
TEGR is the only such branch; the rank-reduced scalar and tensor-only
branches must also be retained.
\endgroup

\subsection{Newtonian Parameter Surface}

Using the Newtonian parameterization in Eq.~\eqref{eq:epsilon},
\begin{equation}
s
=
\frac{\epsilon}
{(\epsilon-1)(4\epsilon-1)},
\qquad
\frac{a_1a_2}{s}
=
-\frac{1}{9\epsilon}.
\label{eq:newtonians}
\end{equation}
The tensor and first-scalar kinetic conditions require
\begin{equation}
0<\epsilon<\frac14,
\label{eq:epsilonrange}
\end{equation}
before imposing the remaining scalar, axial, and vector conditions. The GR point $\epsilon=0$ lies on the rank-changing surface $s=0$ and must be obtained from the unreduced theory.

\begin{table}[t]
\centering
\caption{Linear behavior on important special parameter surfaces.}
\label{tab:speciallimits}
\renewcommand{\arraystretch}{1.3}
\setlength{\tabcolsep}{8pt}
\begin{tabular}{lll}
\hline
\textbf{Condition}
&
\textbf{Immediate consequence}
&
\textbf{Required treatment}
\tabularnewline
\hline
\(a_4=0\), generic
&
Parity-even theory; equal vector speeds
&
Check gradient stability
\tabularnewline
\hline
\(s=0\), \(a_4\neq0\)
&
One scalar combination remains
&
Use Eq.~\eqref{eq:szeroconstraints}
\tabularnewline
\hline
\(f_1=0\)
&
Pseudoscalar kinetic term vanishes
&
Study nonlinear strong coupling
\tabularnewline
\hline
\(a_4^2=M_3^2\)
&
One vector kinetic eigenvalue vanishes
&
Reduced-rank analysis
\tabularnewline
\hline
\(c=0\), \(a_4\neq0\)
&
Nonhyperbolic vector equation
&
Exclude from the wave region
\tabularnewline
\hline
{\(c=0\), \(a_4=0\)}
&
{Tensor plus scalar, generically}
&
{Analyze the rank-reduced action}
\tabularnewline
\hline
{\(c=0\), \(a_4=0\), \(a_2=0\)}
&
{Only two tensor modes at linear order}
&
{Check nonlinear strong coupling}
\tabularnewline
\hline
TEGR
&
Only two tensor modes remain
&
Use the unreduced action
\tabularnewline
\hline
\end{tabular}
\end{table}

\section{Representative Reduced-Action Parameter Scan}
\label{7}

We illustrate the kinetic and propagation conditions of the generic reduced theory using a representative point on the Newtonian parameter surface:
\begin{equation}
\epsilon=0.05,
\qquad
a_3=0.
\label{eq:benchmarkinput}
\end{equation}
Using Eqs.~\eqref{eq:epsilon} and \eqref{eq:newtonians}, we obtain
\begin{equation}
a_1=-0.350877,
\qquad
a_2=0.416667,
\qquad
s=0.0657895,
\qquad
c=-0.350877.
\label{eq:benchmarkparameters}
\end{equation}

The relevant bounds on the parity-odd coupling are
\begin{equation}
M_1^2=0.207756,
\qquad
M_2^2=4.15512,
\qquad
M_3^2=0.0890384.
\label{eq:benchmarkbounds}
\end{equation}
The first kinetic-degeneracy boundary is therefore determined by $M_3$:
\begin{equation}
|a_4|
<
\sqrt{M_3^2}
=
0.298393.
\label{eq:benchmarkinterval}
\end{equation}

Within this interval, the tensor and scalar kinetic terms are positive, both vector kinetic eigenvalues are positive, the vector characteristic discriminant is nonnegative, and both squared vector speeds are positive. Thus, the system is kinetically ghost-free and has hyperbolic vector propagation.

However, these conditions do not imply complete stability. Throughout the same interval,
\begin{equation}
\operatorname{tr}\bm G_V
=
-\operatorname{tr}\bm M_V
<
0.
\end{equation}
Therefore, at least one vector gradient-energy eigenvalue is negative. The vector Hamiltonian is unbounded from below, so the generic theory is not stable even though its kinetic eigenvalues are positive and its propagation speeds are real.

\begin{figure}[t]
\centering
\IfFileExists{vector_parameter_scan.pdf}{%
  \includegraphics[width=\columnwidth]{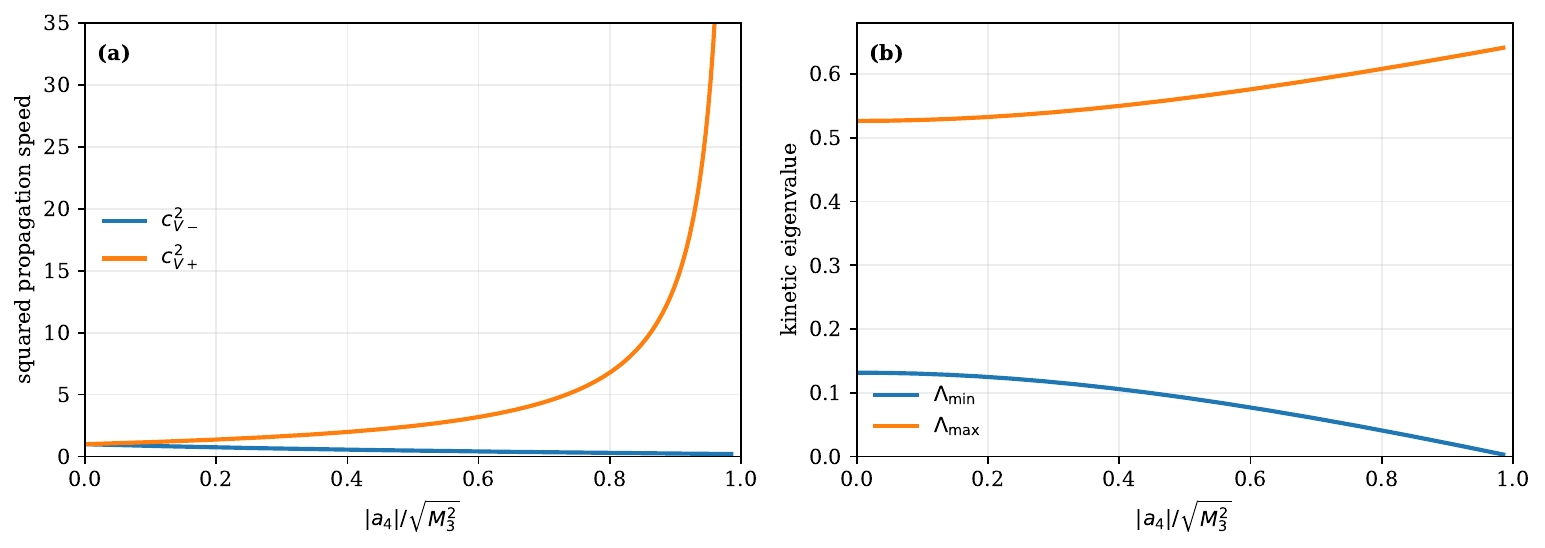}%
}{%
  \IfFileExists{figures/vector_parameter_scan.pdf}{%
    \includegraphics[width=\columnwidth]{vector_parameter_scan.pdf}%
  }{%
    \fbox{\parbox{0.92\columnwidth}{\centering
    {The file \texttt{vector\_parameter\_scan.pdf}
    must be placed in the manuscript directory or in the
    \texttt{figures} subdirectory.}}}%
  }%
}
\caption{Representative vector-sector scan for $\epsilon=0.05$ and $a_3=0$. (a) Squared propagation speeds of the two vector branches. (b) Eigenvalues of the vector kinetic matrix. The dashed line indicates $|a_4|/\sqrt{M_3^2}=1$, where one kinetic eigenvalue vanishes. Although the displayed interval has positive kinetic eigenvalues, it is not a stable energy-bounded region because the vector Hamiltonian is unbounded; see Sec.~\ref{sec:vector-boundedness}.}
\label{fig:vectorscan}
\end{figure}

Figure~\ref{fig:vectorscan} shows the squared vector speeds and vector kinetic eigenvalues as functions of the normalized parity-odd coupling. At $a_4=0$, both vector branches propagate at the speed of light. As $|a_4|$ increases, the two speeds split: one branch becomes subluminal, while the other becomes superluminal and diverges near the kinetic-degeneracy boundary. At the same time, the smallest kinetic eigenvalue, $\Lambda_{\min}$, approaches zero. Thus, the divergent propagation speed signals kinetic degeneracy rather than a physically regular large-speed regime.
Several representative values are listed in Table~\ref{tab:vectorscan}. The vector branches are labeled such that
\begin{equation}
c_{V-}\leq c_{V+}, \quad \mbox{for $a_4>0$.}
\end{equation}

\begin{table}[t]
\centering
\caption{Vector propagation speeds and kinetic eigenvalues for the
representative parameter set. Here,
\(x=|a_4|/\sqrt{M_3^2}\).}
\label{tab:vectorscan}
\renewcommand{\arraystretch}{1.25}
\setlength{\tabcolsep}{8pt}
\begin{tabular}{cccccc}
\hline
\(\bm{x}\) &
\(\bm{|a_4|}\) &
\(\bm{c_{V-}}\) &
\(\bm{c_{V+}}\) &
\(\bm{\Lambda_{\min}}\) &
\(\bm{\Lambda_{\max}}\)
\tabularnewline
\hline
0.00 & 0.000000 & 1.000 & 1.000 & 0.13158 & 0.52632
\tabularnewline
\hline
0.25 & 0.074598 & {blue}{0.908} & {blue}{1.122} & 0.12116 & 0.53585
\tabularnewline
\hline
0.50 & 0.149197 & {blue}{0.830} & {blue}{1.314} & 0.09243 & 0.56194
\tabularnewline
\hline
0.75 & 0.223795 & {blue}{0.756} & {blue}{1.742} & 0.05055 & 0.59942
\tabularnewline
\hline
0.90 & 0.268554 & {blue}{0.710} & {blue}{2.613} & 0.02104 & 0.62544
\tabularnewline
\hline
\end{tabular}
\end{table}
This example shows that the kinetic-positivity and hyperbolicity conditions have a nonempty parameter region. However, the identity
\begin{equation}
\operatorname{tr}\bm G_V
=
-\operatorname{tr}\bm M_V
\end{equation}
shows that the vector Hamiltonian remains unbounded in this region. The scan therefore illustrates the difference between real propagation speeds and energetic stability; it does not identify a new stable branch. This result agrees with the gauge-invariant analysis of Ref.~\cite{BahamondeEtAl2025}.

\section{Full Sourced SVT Response}
\label{sec:full-sourced-svt}

We now couple the linear tetrad perturbations to a general compact canonical source. We use the interaction convention
\begin{equation}
S_{\rm int}^{(1)}
=
\int\dd^4x\,a_{\mu\nu}T^{\mu\nu}
=
\int\dd^4x
\left[
\frac12h_{\mu\nu}T^{(\mu\nu)}
+
A_{\mu\nu}T^{[\mu\nu]}
\right].
\label{eq:linear-source-coupling}
\end{equation}
Changing the overall sign convention changes the signs of the field amplitudes, but not the quadratic energy fluxes.

For each nonzero Fourier mode, the source is decomposed into scalar, vector, tensor, and antisymmetric components:
\begin{align}
T^{00}&=\rho,
\quad
T^{(0i)}=\partial_iq+q_i,
\quad
T^{[0i]}=\partial_i\sigma+\sigma_i,
\nonumber\\
T^{(ij)}
&=
p\,\delta_{ij}
+
\left(
\partial_i\partial_j-\frac13\delta_{ij}\Delta
\right)\pi
+
2\partial_{(i}\pi_{j)}
+
\Pi_{ij}^{\TT},
&
T^{[ij]}
&=
\epsilon_{ijk}
\left(
\tau_k+\partial_k\tau
\right).
\label{eq:sourceSVT}
\end{align}
The transverse variables satisfy
\begin{equation}
\partial_iq_i
=
\partial_i\sigma_i
=
\partial_i\pi_i
=
\partial_i\tau_i
=
0,
\qquad
\partial_i\Pi_{ij}^{\TT}
=
0,
\qquad
\Pi_{ii}^{\TT}=0.
\end{equation}

Conservation of the canonical translation current,
\begin{align}
\partial_\mu T^{\mu\nu}=0, \quad \mbox{implies} \quad
\dot\rho+\Delta(q-\sigma)=0,
\quad
\dot q+\dot\sigma+p+\frac23\Delta\pi=0,
\quad
\dot q_i+\dot\sigma_i+\Delta\pi_i
-(\bm\nabla\times\bm\tau)_i=0.
\end{align}
These relations show that not all source components are independent radiative data.

In the scalar gauge, the source projections are
\begin{equation}
j_\phi=\rho,
\qquad
j_\psi=3p,
\qquad
j_\mu=-2\Delta\sigma,
\qquad
j_\chi=-2\Delta\tau.
\label{eq:scalar-source-projections}
\end{equation}
The transverse vector projections are
\begin{equation}
j_{\beta i}=-(q_i+\sigma_i),
\quad
j_{\gamma i}=q_i-\sigma_i,
\quad
j_{\lambda i}=2\tau_i.
\label{eq:vector-source-projections}
\end{equation}
The tensor projection is
\begin{equation}
j_{TA}
=
\frac12e_A^{ij}\Pi_{ij}^{\TT}.
\label{eq:tensor-source-projection}
\end{equation}
\begingroup
With \(d=a_1-2a_2\), the two sourced scalar constraints are
\begin{align}
\phi={}\dot\mu-\frac{d}{s}\psi
-\frac{a_4}{3s}\dot\chi
-\frac{\kappa}{2sk^2}j_\phi,
\qquad
\mu={}
\frac{9a_2\dot\psi+a_4k^2\chi}{3sk^2}
+\frac{\kappa}{2sk^4}
\left(j_\mu-\dot j_\phi\right).
\label{eq:sourced-mu-constraint}
\end{align}
These formulas apply to \(k>0\); the zero-momentum charges belong to
the stationary sector and must be treated separately.
\endgroup
After solving the scalar constraints, the propagating scalar fields satisfy
\begin{align}
\frac{18a_1a_2}{s}
\left(
\ddot\psi+k^2\psi
\right)
=
-\kappa J_\psi^{\rm eff},
\qquad
\frac{2f_1}{s}k^2
\left(
\ddot\chi+k^2\chi
\right)
=
-\kappa J_\chi^{\rm eff},
\end{align}
where
\begin{align}
J_\psi^{\rm eff}
=
j_\psi
-\frac{a_1-2a_2}{s}j_\phi
-\frac{3a_2}{sk^2}
\left(
\dot j_\mu-\ddot j_\phi
\right),
\qquad
J_\chi^{\rm eff}
=
j_\chi+\frac{a_4}{3s}j_\mu.
\end{align}
The tensor modes satisfy
\begin{equation}
{\frac{3a_1}{2}}
\left(
\ddot H_A+k^2H_A
\right)
=
-\kappa j_{TA}.
\label{eq:sourced-tensor}
\end{equation}
{The factor \(1/2\) is fixed by
\(S_T^{(2)}=-3a_1\int\dd^4x\,
[\dot h_{ij}^{\TT}\dot h_{ij}^{\TT}
-\partial_kh_{ij}^{\TT}\partial_kh_{ij}^{\TT}]/(8\kappa)\)
and by \(e_A^{ij}e_{Bij}=2\delta_{AB}\).  It also reproduces the
standard linearized Einstein equation at the TEGR point.}

For the vector sector, eliminating the nondynamical variable $\gamma_A$ gives
\begingroup
\begin{align}
-4c_1k^2\gamma_A+2c_2k^2\beta_A
&+2(c_2-2c_1)p_Ak\dot\lambda_A-c_4k^2\lambda_A
-c_4p_Ak\dot\beta_A
=2\kappa j_{\gamma A},
\label{eq:explicit-sourced-gamma}\\
\gamma_A&=
\frac{2c_2k^2\beta_A
+2(c_2-2c_1)p_Ak\dot\lambda_A-c_4k^2\lambda_A
-c_4p_Ak\dot\beta_A-2\kappa j_{\gamma A}}
{4c_1k^2}.
\label{eq:explicit-sourced-gamma-solution}
\end{align}
\endgroup
\noindent
Substitution into the remaining two vector equations gives
\begin{equation}
K_{V,A}^{\rm eff}q_A
=
-\kappa j_{V,A}^{\rm eff},
\end{equation}
where
\begin{equation}
K_{V,A}^{\rm eff}
=
K_{qq}-K_{qn}K_{nn}^{-1}K_{nq},
\qquad
j_{V,A}^{\rm eff}
=
j_q-K_{qn}K_{nn}^{-1}j_n.
\label{eq:vector-Schur}
\end{equation}
The first expression is the Schur complement of the vector operator, and the second gives the effective source for the two propagating vector modes.

For any reduced Fourier operator $K_R(\omega,k)$, the retarded solution is
\begin{equation}
Q_R(x)
=
-\kappa
\int\dd^4x'\,
G_R^{\rm ret}(x-x')
j_R^{\rm eff}(x'),
\end{equation}
with
\begin{equation}
G_R^{\rm ret}(\omega,\bm k)
=
\left[
K_R(\omega,k)+i0\,\omega
\right]^{-1}.
\label{eq:full-retarded-matrix}
\end{equation}
The far-zone amplitudes are obtained from the outgoing poles of this Green function. This gives explicit source-to-amplitude relations for the two tensor, two scalar, and four vector polarizations. The vector responses are formal signed solutions in the generic branch because the vector Hamiltonian is unbounded.
\begingroup
More explicitly, for a simple outgoing pole
\(k=k_s(\omega)=\omega/c_s\), introduce right and left null vectors
\begin{equation}
K_R(\omega,k_s)e_s=0,
\qquad
\ell_s^\dagger K_R(\omega,k_s)=0.
\end{equation}
Radial residue evaluation gives
\begin{align}
Q_R(\omega,r\bm n)
&=-\frac{\kappa}{4\pi r}
\sum_s e^{ik_sr}\,
\mathcal R_s(\omega)
j_R^{\rm eff}(\omega,k_s\bm n)
+O(r^{-2}),
\label{eq:source-far-zone}\\
\mathcal R_s(\omega)
&=
\frac{2k_s e_s\ell_s^\dagger}
{\ell_s^\dagger
[\partial_kK_R(\omega,k)]_{k_s}e_s}.
\label{eq:spatial-pole-residue}
\end{align}
This expression assumes a simple pole and a localized source; a
multiple pole or a rank-changing surface must be treated separately.
For the tensor and scalar blocks it yields
\begin{align}
H_A(\omega,\bm n)
&=-\frac{\kappa}{6\pi a_1}\,
j_{TA}(\omega,\omega\bm n),
\label{eq:tensor-source-amplitude}\\
\psi_0(\omega,\bm n)
&=-\frac{\kappa s}{72\pi a_1a_2}\,
J_\psi^{\rm eff}(\omega,\omega\bm n),
\label{eq:psi-source-amplitude}\\
\chi_0(\omega,\bm n)
&=-\frac{\kappa s}{8\pi f_1\omega^2}\,
J_\chi^{\rm eff}(\omega,\omega\bm n).
\label{eq:chi-source-amplitude}
\end{align}
For the vector branch \(s=(\sigma,A)\),
\begin{equation}
F_{\sigma A}(\omega,\bm n)e_\sigma^{(A)}
=-\frac{\kappa}{4\pi}
\mathcal R_{\sigma A}(\omega)
j_{V,A}^{\rm eff}
\left(\omega,\frac{\omega}{c_{V\sigma}}\bm n\right).
\label{eq:vector-source-amplitude}
\end{equation}
Equations~\eqref{eq:chi-source-amplitude} and \eqref{eq:vector-source-amplitude}
are the source-to-amplitude map claimed in the abstract.  They determine
linear far-zone amplitudes as functionals of a specified conserved
source; they are not, by themselves, a near-zone model of a compact
binary.
\endgroup

\section{Spin-Current Example and Outgoing-Mode Parametrization}
\label{8}

Section~\ref{sec:full-sourced-svt} determines the far-zone amplitudes of all linear modes from the SVT components of a general compact source. Here, we write the outgoing-wave form of these modes and use the composite field
\begin{equation}
V_\mu
=
\partial^\nu B_{\nu\mu},
\end{equation}
as an example of a field sourced by intrinsic spin. This field provides useful information about the antisymmetric sector, but it does not determine the full tensor $B_{\mu\nu}$.

For a mode with speed $c_N$, define the retarded time
\begin{equation}
u_N
=
t-\frac{r}{c_N}.
\end{equation}
For luminal tensor and scalar modes, $u=t-r$. For the two vector branches,
\begin{equation}
u_\pm
=
t-\frac{r}{c_{V\pm}}.
\end{equation}

The far-zone tetrad perturbation can be written as
\begin{align}
a_{\mu\nu}^{\rm rad}
={}&
{\frac1{2r}}
\sum_{A=+,\times}
H_A(u,\bm n)e_{\mu\nu}^A
+
\frac1r
\psi_0(u,\bm n)e_{\mu\nu}^{(S)}
+
\frac1r
\chi_0(u,\bm n)e_{\mu\nu}^{(P)}
\nonumber\\
&+
\frac1r
\sum_{\lambda=L,R}
\left[
V_{+,\lambda}(u_+,\bm n)e_{\mu\nu}^{(V+,\lambda)}
+
V_{-,\lambda}(u_-,\bm n)e_{\mu\nu}^{(V-,\lambda)}
\right]
+
O(r^{-2}).
\label{eq:outgoing}
\end{align}
{The factor \(1/2\) in the tensor term follows from
\(a_{(\mu\nu)}=h_{\mu\nu}/2\); \(H_A\) is the metric tensor amplitude
used in Eqs.~\eqref{eq:sourced-tensor} and \eqref{eq:tensorflux}.}
The tensor, scalar, pseudoscalar, and vector polarization tensors are determined by the SVT eigenvectors and normalized by their quadratic actions. The vector amplitudes $V_{\pm,\lambda}$ should not be confused with the composite field $V_\mu$.

For a field of the form
\begin{equation}
\Phi_N
=
\frac{F_N(u_N,\bm n)}{r},
\end{equation}
the leading derivatives satisfy
\begin{equation}
\partial_t\Phi_N
=
\frac{\dot F_N}{r}
+
O(r^{-2}),
\qquad
\partial_r\Phi_N
=
-\frac{\dot F_N}{c_Nr}
-
\frac{F_N}{r^2}
+
O(r^{-2}),
\qquad
\left(
\partial_t^2-c_N^2\Delta
\right)\Phi_N
=
O(r^{-3}).
\end{equation}
Thus, the leading radiation field decays as $1/r$.

\subsection{Spin-Current Source}

The retarded solution of Eq.~\eqref{eq:Vsourced} is
\begin{equation}
V_\mu(t,\bm x)
=
-\frac{\kappa}{4\pi}
\int
\frac{
\partial'^\nu T_{[\mu\nu]}
\left(
t-|\bm x-\bm x'|,
\bm x'
\right)
}{
|\bm x-\bm x'|
}
\dd^3x'.
\label{eq:retardedV}
\end{equation}
For an observer far from a compact source,
\begin{equation}
|\bm x-\bm x'|
=
r-\bm n\cdot\bm x'
+
O(r^{-1}),
\end{equation}
and
\begin{equation}
V_\mu(t,\bm x)
=
-\frac{\kappa}{4\pi r}
\int\dd^3x'\,
\partial'^\nu T_{[\mu\nu]}
\left(
u+\bm n\cdot\bm x',
\bm x'
\right)
+
O(r^{-2}).
\label{eq:farV}
\end{equation}

Define the conserved effective antisymmetric source
\begin{equation}
J_\mu
=
\partial^\nu T_{[\mu\nu]},
\qquad
\partial^\mu J_\mu=0.
\label{eq:Jdefinition}
\end{equation}
Its multipole moments are
\begin{equation}
\mathcal J_\mu^L(u)
=
\int\dd^3x'\,
x'^LJ_\mu(u,\bm x').
\label{eq:Jmoments}
\end{equation}
The far-zone expansion becomes
\begin{equation}
V_\mu(t,r\bm n)
=
-\frac{\kappa}{4\pi r}
\sum_{\ell=0}^\infty
\frac{1}{\ell!}
n_L
\frac{\dd^\ell}{\dd u^\ell}
\mathcal J_\mu^L(u)
+
O(r^{-2}).
\label{eq:Vmultipoles}
\end{equation}

\subsection{Intrinsic-Spin Multipoles}

For matter with intrinsic spin, the antisymmetric energy-momentum tensor satisfies the Tetrode relation
\begin{equation}
T_{[\mu\nu]}
=
\frac12
\partial_\lambda
S_{\mu\nu}{}^\lambda,
\end{equation}
where $S_{\mu\nu}{}^\lambda$ is the canonical spin current. Therefore,
\begin{equation}
J_\mu
=
\frac12
\partial^\nu\partial_\lambda
S_{\mu\nu}{}^\lambda.
\label{eq:Jspin}
\end{equation}

The spin multipoles are
\begin{equation}
\mathcal S_{\mu\nu}^{\lambda|L}(u)
=
\int\dd^3x\,
x^L
S_{\mu\nu}{}^\lambda(u,\bm x).
\label{eq:spinmoments}
\end{equation}
For a compact source, integration by parts gives
\begin{equation}
\mathcal J_\mu
=
-\frac12
\ddot{\mathcal S}_{\mu0}^{0}.
\label{eq:Jmonopolespin}
\end{equation}
Since $S_{00}{}^\lambda=0$, one has
\begin{equation}
\mathcal J_0=0.
\end{equation}

The dipole source moment is
\begin{equation}
\mathcal J_\mu^a
=
-\frac12
\ddot{\mathcal S}_{\mu0}^{0|a}
+
\frac12
\dot{\mathcal S}_{\mu0}^{a}
-
\frac12
\dot{\mathcal S}_{\mu a}^{0}.
\label{eq:Jdipolespin}
\end{equation}
The ordinary total spatial spin is
\begin{equation}
\Sigma^b
=
\frac12
\epsilon^{bij}
\mathcal S_{ij}^{0}.
\label{eq:totalspin}
\end{equation}
Its contribution to the spatial radiation field is
\begin{equation}
\left.V_i\right|_{\Sigma}
=
\frac{\kappa}{8\pi r}
n_a\epsilon_{iab}
\ddot\Sigma^b
+
O(r^{-2}).
\label{eq:spinradiativeV}
\end{equation}
Thus, a constant total intrinsic spin does not radiate in this channel. The corresponding field strength contains an additional time derivative and is proportional to $\dddot\Sigma^i/r$.

\subsection{Physical Interpretation}

The spin-current example leads to several conclusions:
\begin{enumerate}
\item A stationary antisymmetric source produces no radiation.
\item The monopole contribution to $V_i$ depends on boost-spin moments, not on ordinary spatial spin.
\item Ordinary intrinsic spin first contributes through the dipole term.
\item A conserved total spin does not radiate through Eq.~\eqref{eq:spinradiativeV}, although higher spin multipoles may radiate.
\item Orbital angular momentum cannot replace the intrinsic-spin current without specifying a consistent matter model.
\end{enumerate}

If the source is described by a symmetric Belinfante energy-momentum tensor, then
\begin{equation}
T_{[\mu\nu]}=0,
\end{equation}
and there is no direct source for $V_\mu$. A prediction for a spinning compact object therefore requires a matter model with a canonical spin current or an equivalent effective antisymmetric source.

\section{Energy--Momentum Balance}
\label{9}

We now relate the outgoing waves to the total energy and momentum of the system. The M{\o}ller energy--momentum complex satisfies
\begin{equation}
\sqrt{-g}
\left(
T_\mu{}^\nu+t_\mu{}^\nu
\right)
=
\partial_\lambda U_\mu{}^{\nu\lambda},
\qquad
U_\mu{}^{\nu\lambda}
=
-U_\mu{}^{\lambda\nu},
\end{equation}
where $U_\mu{}^{\nu\lambda}$ is the M{\o}ller superpotential, $T_\mu{}^\nu$ is the matter energy--momentum tensor, and $t_\mu{}^\nu$ is the gravitational energy--momentum complex. Therefore,
\begin{equation}
\partial_\nu
\left[
\sqrt{-g}
\left(
T_\mu{}^\nu+t_\mu{}^\nu
\right)
\right]
=
0.
\end{equation}

Outside a compact source, $T_\mu{}^\nu=0$. The energy--momentum balance law at infinity is then
\begin{equation}
\frac{\dd P_\mu}{\dd t}
=
-\lim_{r\rightarrow\infty}
\oint_{S_r}
\sqrt{-g}\,
t_\mu{}^\alpha
n_\alpha r^2\dd\Omega.
\label{eq:balance}
\end{equation}
The radiative flux is second order because the first-order torsion decays as $1/r$, while the quadratic gravitational current decays as $1/r^2$.

For a diagonal mode with amplitude $\Phi_N$, speed $c_N$, and signed normalization $Z_N$, the quadratic action is
\begin{equation}
S_N^{(2)}
=
\frac12
\int\dd^4x\,
Z_N
\left[
\dot\Phi_N^2
-
c_N^2(\bm\nabla\Phi_N)^2
\right].
\end{equation}
An outgoing wave,
\begin{equation}
\Phi_N
=
\frac{F_N(u_N,\bm n)}{r}
+
O(r^{-2}), \qquad \mbox{carries the flux} \quad
\frac{\dd E_N}{\dd t\,\dd\Omega}
=
Z_Nc_N\dot F_N^2.
\label{eq:modeflux}
\end{equation}
Positive outgoing energy requires $Z_N>0$. The total mass-loss relation is
\begin{equation}
\frac{\dd M}{\dd t}
=
-\int\dd\Omega
\sum_N
Z_Nc_N\dot F_N^2.
\label{eq:massloss}
\end{equation}
\begingroup
For the vector eigenmodes used below, the normalization entering
Eq.~\eqref{eq:modeflux} is $Z_N=Z_{V\sigma}^{(A)}/\kappa$; the
dimensionless quantity $Z_{V\sigma}^{(A)}$ is fixed by the eigenvector
normalization in Eq.~\eqref{eq:vectorfluxexplicit}.
\endgroup

\subsection{M{\o}ller--Noether Identity}

The mode fluxes must agree with the current that defines the M{\o}ller energy--momentum. Let $\mathscr L_G$ be the full gravitational Lagrangian density and define the tetrad momentum
\begin{equation}
\Pi_k{}^{\rho\nu}
=
\frac{\partial\mathscr L_G}
{\partial(\partial_\nu b^k{}_\rho)}.
\label{eq:tetradmomentum}
\end{equation}
The M{\o}ller superpotential is
\begin{equation}
U_\mu{}^{\nu\lambda}
=
b^k{}_\mu\Pi_k{}^{\nu\lambda},\quad \mbox{while the canonical gravitational current is} \quad
\mathfrak t_\mu{}^\nu
=
\Pi_k{}^{\rho\nu}
\partial_\mu b^k{}_\rho
-
\delta_\mu{}^\nu\mathscr L_G.
\label{eq:fullcanonicalcomplex}
\end{equation}

The tetrad field equations imply
\begin{equation}
\partial_\lambda U_\mu{}^{\nu\lambda}
=
\sqrt{-g}\,T_\mu{}^\nu
+
\mathfrak t_\mu{}^\nu.
\label{eq:exactmolleridentity}
\end{equation}
Expanding around Minkowski spacetime, the gravitational Lagrangian begins at quadratic order. The second-order M{\o}ller current is therefore
\begin{equation}
\mathfrak t_\mu{}^{\nu(2)}
=
\frac{
\partial\mathcal L_G^{(2)}
}{
\partial(\partial_\nu a_{\rho\sigma})
}
\partial_\mu a_{\rho\sigma}
-
\delta_\mu{}^\nu\mathcal L_G^{(2)}
\equiv
\Theta_\mu{}^\nu.
\label{eq:pointwiseMollerNoether}
\end{equation}
\begingroup
Thus, for the first-derivative tetrad Lagrangian used here and with the
canonical current defined in Eq.~\eqref{eq:fullcanonicalcomplex}, the
second-order M{\o}ller current agrees on shell with the Noether current of
the quadratic action. This is a pointwise identity in the present
Lagrangian and tetrad convention. Adding a total divergence to the
Lagrangian changes the local representative by the divergence of an
antisymmetric superpotential; under the stated wave-zone falloff, this
does not change the sphere-integrated charge or radiative flux. The
physically relevant statement is therefore the equality of the
surface-integrated M{\o}ller and Noether fluxes, rather than an
improvement-independent uniqueness of either local current.
\endgroup

\subsection{Energy and Momentum Fluxes}

The tensor energy flux is
\begin{equation}
\frac{\dd E_T}{\dd t\,\dd\Omega}
=
-\frac{3a_1}{2\kappa}
\left\langle
\dot H_+^2+\dot H_\times^2
\right\rangle.
\label{eq:tensorflux}
\end{equation}
For $a_1=-1/3$, this reduces to the standard general-relativistic normalization.

For the scalar modes, define
\begin{equation}
X
=
\sqrt{-\Delta}\,\chi.
\label{eq:Xdefinition}
\end{equation}
The scalar flux is
\begin{equation}
\frac{\dd E_S}{\dd t\,\dd\Omega}
=
-\frac{18a_1a_2}{\kappa s}
\left\langle
\dot\psi_0^2
\right\rangle
-
\frac{2f_1}{\kappa s}
\left\langle
\dot X_0^2
\right\rangle.
\label{eq:scalarfluxexplicit}
\end{equation}

The total vector flux is
\begin{equation}
\frac{\dd E_V}{\dd t\,\dd\Omega}
=
\frac1\kappa
\sum_{\sigma=\pm}
\sum_{A=L,R}
Z_{V\sigma}^{(A)}
c_{V\sigma}
\left|
\dot F_{\sigma A}
\right|^2.
\label{eq:totalvectorflux}
\end{equation}

The complete radiated energy flux is
\begin{align}
\mathcal F_E(\bm n)
\equiv
\frac{\dd E_{\rm rad}}{\dd t\,\dd\Omega}
={}
-\frac{3a_1}{2\kappa}
\left\langle
\dot H_+^2+\dot H_\times^2
\right\rangle
-
\frac{18a_1a_2}{\kappa s}
\left\langle
\dot\psi_0^2
\right\rangle
-
\frac{2f_1}{\kappa s}
\left\langle
\dot X_0^2
\right\rangle
+
\frac1\kappa
\sum_{\sigma=\pm}
\sum_{A=L,R}
Z_{V\sigma}^{(A)}
c_{V\sigma}
\left|
\dot F_{\sigma A}
\right|^2.
\label{eq:completeenergyflux}
\end{align}
The tensor and scalar terms are nonnegative when their no-ghost conditions hold. The vector contribution remains signed because the generic vector Hamiltonian is unbounded. Therefore, Eq.~\eqref{eq:completeenergyflux} is a formal balance law, not a positive-definite mass-loss theorem.

\begingroup
After substituting Eqs.~\eqref{eq:tensor-source-amplitude}--\eqref{eq:vector-source-amplitude},
Eq.~\eqref{eq:completeenergyflux} becomes an explicit quadratic functional
of the conserved SVT source projections. Its numerical value is not fixed
until a matter model and near-zone solution specify those projections.
\endgroup

The total energy loss is
\begin{equation}
\frac{\dd M}{\dd t}
=
-\int\dd\Omega\,
\mathcal F_E(\bm n).
\label{eq:explicitmassloss}
\end{equation}
The corresponding momentum flux is
\begin{equation}
\frac{\dd P_i}{\dd t}
=
-\int\dd\Omega\,n_i
\left[
\mathcal F_{TS}
+
\sum_{\sigma=\pm}
\sum_{A=L,R}
\frac{
\mathcal F_{V\sigma A}
}{
c_{V\sigma}
}
\right].
\label{eq:momentumloss}
\end{equation}

These expressions give the energy and momentum fluxes for any compact canonical source satisfying the assumed conservation laws. Numerical waveforms for a binary or spinning object require a specific matter model that determines the source projections.

\section{Stationary Limit}
\label{10}

When all retarded-time derivatives vanish, the radiative fields carry no energy or momentum flux. The asymptotic solution then reduces to the stationary result obtained in Ref.~\cite{ShirafujiNashed1997}. The first-order surface integrals give
\begin{equation}
E=m,
\qquad
P_\alpha=\mathring B_\alpha.
\end{equation}

For an isolated stationary system in its rest frame,
\begin{equation}
P_\alpha=0,
\end{equation}
and therefore
\begin{equation}
\mathring B_\alpha=0.
\end{equation}
The time-dependent analysis extends this stationary result. The change in the source four-momentum is equal to minus the four-momentum carried away by the outgoing tetrad modes.

\section{Discussion and Conclusion}
\label{11}

We investigated the time-dependent asymptotic sector of the four-parameter tetrad theory introduced in Ref.~\cite{ShirafujiNashed1997}. Starting from the quadratic torsion action, we derived the linearized field equations without assuming stationarity. The divergence of the coupled antisymmetric field satisfies a massless wave equation, and the scalar--vector--tensor decomposition separates the constraints from the propagating fields. In the generic nondegenerate parity-violating region, the Minkowski spectrum contains two tensor, two scalar, and four vector degrees of freedom.

The tensor and scalar modes propagate at the speed of light on Minkowski spacetime. The vector sector contains two branches with propagation speeds controlled by the parity-odd coupling. The two helicities of each branch remain degenerate on a torsion-free Minkowski background. The parameter transformation to the modern invariant basis connects our flat-spacetime conditions with recent gauge-invariant stability analyses.

The special parameter surfaces have distinct physical meanings. At the TEGR point, the antisymmetric tetrad perturbation becomes a Lorentz-gauge variable, leaving only the two tensor polarizations. On the surface $s=0$, the scalar constraint structure changes. The surfaces $f_1=0$ and $a_4^2=M_3^2$ correspond to vanishing scalar and vector kinetic terms, respectively. In the generic branch, the identity
\begin{equation}
\operatorname{tr}\bm G_V
=
-\operatorname{tr}\bm M_V
\end{equation}
proves that the vector Hamiltonian is unbounded from below. Therefore, positive vector kinetic eigenvalues and real propagation speeds do not define a stable region. This result agrees with the parity-even gauge-invariant analysis of Ref.~\cite{BahamondeEtAl2025}.

We also derived the source-generated outgoing response for all propagating modes. A complete SVT decomposition of a canonical source determines the tensor and scalar amplitudes after solving the scalar constraints. The vector amplitudes are obtained from the exact Schur complement and the pole residues of the reduced vector operator. The composite field
\begin{equation}
V_\mu
=
\partial^\nu B_{\nu\mu}
\end{equation}
provides an example of a spin-sourced antisymmetric response. The Tetrode relation shows that ordinary intrinsic spin first contributes at dipole order, and a constant total intrinsic spin does not radiate through this particular field. Although $V_\mu$ does not determine the full antisymmetric tetrad perturbation, the complete source-to-mode map does not depend on such an inversion.

The M{\o}ller superpotential provides a unified description of stationary charges and radiative energy--momentum balance. At first order, it reproduces
\begin{equation}
E=m,
\qquad
P_\alpha=\mathring B_\alpha.
\end{equation}
At second order, the canonical tetrad momentum shows that the quadratic Noether current equals the second-order M{\o}ller current pointwise in the vacuum wave zone. This gives positive energy fluxes for the tensor and scalar sectors when their kinetic conditions are satisfied. The vector contribution remains signed and indefinite because the generic vector Hamiltonian is unbounded. The resulting expression is therefore an energy--momentum balance law, not a proof of stability.

In summary, we derived a complete linear source-to-amplitude map for the eight-mode spectrum, connected these amplitudes to the quadratic M{\o}ller energy--momentum current, and obtained a retarded multipole expansion for the spin-sourced composite field. The radiated luminosity and recoil are explicit functionals of the SVT source projections. Their numerical evaluation for a specific binary or compact object requires a consistent matter and spin model, together with near-zone matching. Future work may study such source models and investigate nonlinear dynamics near the kinetic-degeneracy surfaces.

\end{document}